\documentclass[12pt,a4paper,final]{iopart}

\usepackage{iopams}  
\usepackage{graphicx}
\usepackage[breaklinks=true,colorlinks=true,linkcolor=blue,urlcolor=blue,citecolor=blue]{hyperref}

\begin{document}



\title[\LaTeXe]{First principles many-body calculations of electronic structure and optical properties of SiC nanoribbons}

\author{Naresh Alaal$^{1,2,3}$}
\address{$^1$IITB-Monash Research Academy, CSE Building 2$^{\,nd}$ Floor, IIT Bombay, Mumbai 400076, India}
\address{$^2$Department of Physics, Indian Institute of Technology Bombay, Mumbai 400076, India}
\address{$^3$Department of Materials Engineering, Monash University, Clayton, Victoria 3800, Australia}

\author{Vaideesh Loganathan $^{2}$}
\address{$^2$Department of Physics, Indian Institute of Technology Bombay, Mumbai 400076, India}

\author{Nikhil Medhekar $^{3}$}
\address{$^3$Department of Materials Engineering, Monash University, Clayton, Victoria 3800, Australia}
\author[cor]{Alok Shukla $^{2,*}$}
\address{$^2$Department of Physics, Indian Institute of Technology Bombay, Mumbai 400076, India}
\eads{\mailto{shukla@phy.iitb.ac.in}}


\begin{abstract}

A first principles many-body approach is employed to calculate the
band structure and optical response of nanometer sized ribbons of
SiC. Many-body effects are incorporated using the GW approximation,
and excitonic effects are included using the Bethe-Salpeter equation.
Both unpassivated and hydrogen passivated armchair SiC nanoribbons
are studied. As a consequence of low dimensionality, large quasiparticle
corrections are seen to the Kohn-Sham energy gaps. In both cases quasiparticle
band gaps are increased by up to 2 eV, as compared to their Kohn-Sham
energy values. Inclusion of electron-hole interactions modifies the
absorption spectra significantly, giving rise to strongly bound excitonic
peaks in these systems.The results suggest that hydrogen-passivated armchair
SiC nanoribbons have the potential to be used in optoelectronic devices
operating in the UV-Vis region of the spectrum.  We also compute the formation energies of these nanoribbons as a function of their widths, and conclude that hydrogen-saturated ribbons will be much more stable as compared to the bare ones.
\end{abstract}

\pacs{Insert suggested PACS numbers here.}


\maketitle

\section{Introduction\label{sec:Section-I}}

Successful synthesis of graphene\cite{graphene_synth}, and discovery
of its extraordinary electronic properties such as high electron mobility
and electrical conductivity\cite{gn1,gn2,gn3}, has tremendously enhanced
research interest in the field of low-dimensional systems, in general,
and graphene-like two-dimensional (2D) systems, in particular. One
of the drawbacks of graphene is the fact that it is a zero bandgap
semiconductor, which limits its applications in electronic devices.
Therefore, it is of great interest to discover novel low-dimensional
materials which share the attractive properties of graphene, but,
in addition, also have a finite band gap. One such candidate, monolayer
of hexagonal boron nitride (h-BN) known as 'white graphene',
has been successfully synthesized\cite{bn_synth} using mechanical
exfoliation and chemical vapor deposition techniques, and electrically
it is a large bandgap insulator. Nanotubes and nanoribbons of boron
nitride have been synthesized and studied\cite{bn1,bn2,bn3,bn4}.
Carbon and silicon are the two most interesting elements in group
IV because of their abundance which leads to low costs, and also because
of the fact that their allotropes possess attractive electronic properties.
Like carbon, silicon also forms a honeycomb structure called silicene
which, not only has been studied theoretically\cite{si_t}, but also
has been experimentally synthesized in high vacuum conditions \cite{si1,si2}.
Si, unlike C, prefers the $sp^{3}$ hybridization, therefore, silicene,
in contrast with graphene, has a buckled bilayer structure. Furthermore,
it has not been possible to synthesize silicene in free-standing form,
thus, limiting its potential for device applications.

Just like carbon, it is well known that the bulk SiC has several allotropes, although 
a layered graphitic one has not been experimentally realized so far. However, theoretical calculations have shown that graphitic SiC can be realized by using the technique of ``extreme hole injection''\cite{graphitic_SiC}. Therefore,
it is conceivable that it can also be isolated in a monolayer form, similar
to graphene \cite{sicmix1,sicmix2}. Because of the fact that basic
unit of SiC contains two different atoms, corresponding monolayer
will exhibit a finite band gap. Similarly, it may be possible to synthesize
SiC-based nanostructures such as nanoribbons, nanotubes, and nanodisks
etc \cite{sicmix1,sicmix2}. From the applications point of view,
SiC nanostructures will have the advantage that their integration
into the present Si based technology will be relatively easier. Keeping
this in mind, in this work we undertake a systematic electronic structure
study of SiC nanoribbons (SiCNRs). Next, we briefly review the existing
studies of SiC, and related structures.

Consistent with the intuition, various electronic structure studies
have predicted SiC monolayer to be a semiconductor with a band gap
of 2.5 eV\cite{SiC_mono,ASiCNR_Eg_osc,sic_bare,sic2d_eg,SiX,sic_bulk_gw,SiC_BN_BeO,group4,opt_SiCNT,
tight_bind}. Although, SiC monolayer has not been realized in the laboratory as
yet, the ultra thin graphitic SiC, which exhibits light emitting properties,
has recently been synthesized \cite{SiC2D}. SiC micro ribbons \cite{sicnr_micro},
crystaline 3C - SiC nanobelts \cite{3C_nanobelts}, bicrystalline
SiC nanobelts \cite{bicrystaline_nanobelt}, and crystalline SiC nanoribbons\cite{SiCNR_synth}
have been successfully synthesized. Nanowires of SiC have  been synthesized by reacting 
SiO with carbon nanotubes at different temperatures\cite{SiCNT_synth}. First principles 
calculations have shown that hydrogenation of graphitic SiC can lead to formation of stable SiC 
nanotubes\cite{SiCNT_stable}.

As far as theory is concerned, SiC nanoribbons (SiCNRs) have been
studied extensively using the density-functional theory (DFT) based
first-principles methodology \cite{ASiCNR_Eg_osc,ZSiCNR_DFT,ZSiCNR_Efield,assymetry_H,edge_H_zig,sic_bare,O_S_zig,mod_eg,B_N_SiCNR,half_bare_zig,vac_SiCNR}.
Sun \emph{et al}.\cite{ASiCNR_Eg_osc} studied hydrogen passivated
SiCNRs using Perdew-Wang functional (PW91) under the generalized-gradient
approximation (GGA) exchange correlation scheme. They found that hydrogen
passivated armchair SiCNRs (ASiCNRs) are semiconducting with a direct
bandgap, which oscillates and exhibits a gradual overall increase
with the width. For the nanoribbons studied, they reported Kohn-sham
band gaps in the range 2.3--2.4 eV \cite{ASiCNR_Eg_osc}. Sun \emph{et
al}.\cite{ASiCNR_Eg_osc} also found that unlike their graphene counterparts,
hydrogen-passivated zigzag SiCNRs (ZSiCNRs) have non degenerate spins,
and were near half metallic, with the smaller gap being a few meVs,
while the larger one being close to 0.6 eV. The ferri-magnetic state
was reported to be most stable, with an anti parellel spin orientation
between the two edges \cite{ASiCNR_Eg_osc}. Bekaroglu \emph{et al.,}\cite{sic_bare}
based upon their first-principles DFT-GGA calculations, reported that
bare ASiCNRs have indirect band gaps in the range 1.3--1.7 eV. The
same authors\cite{sic_bare} as well as Morbec \emph{et al.}\cite{vac_SiCNR}
studied the influence of vacancies on the electronic structure of
SiCNRs using the DFT-GGA approach, and found that single (Si and C)
vacancies induce magnetism in non-magnetic ASiCNRs, and double (Si
and C) vacancies induce magnetism in ZSiCNRs. Tang \emph{et al.} \cite{half_bare_zig}
studied bare-edge and half bare-edge ZSiCNRs using the DFT-GGA approach
and reported that the bare-edge ZSiCNRs exhibit magnetic metallic
behavior, while half bare edge ZSiCNRs are magnetic semiconductors
with small band gap. Oxygen and sulfur terminated ZSiCNRs were studied
by Lopez-Benzanilla\emph{ et al}.\cite{O_S_zig} using DFT based local
spin-density approximation (LSDA), and observed that ZSiCNRs can be
tuned from half metallic to semiconducting. The influence of substitution
of edge Si and C atoms with B and N atoms in ZSiCNRs was studied by
Zheng \emph{et al}\cite{mod_eg} using the DFT-GGA approach, and they
found that such substituted ZSiCNRs exhibit semicounducting behavior.
The influence of doping on the electronic and magnetic properties
of SiCNRs have also been studied by Costa \emph{et al}. \cite{B_N_SiCNR},
who showed that the N doped ZSiCNRs are metallic.

Generally speaking nanoribbons of various materials show interesting
properties that are vastly different from those of their bulk counterparts
due to quantum confinement. The bandgap in armchair nanoribbons (ANRs)
oscillates over three families according to their width, and approaches
a value close to that of the parent 2D sheet for large widths \cite{AGNR_Eg_osc,ASiCNR_Eg_osc,ABNNR_Eg_osc}.
Zigzag nanoribbons(ZNRs) of both graphene and SiC are spin polarized,
and the individual spin gaps can be controlled by application of an
electric field directed along the width\cite{ZGNR_Efield,ZSiCNR_Efield,ABNNR_Eg_osc}.
These unique properties of nanoribbons help to tune the band gap on
the basis of their widths, and the edge geometries. Therefore, a good
knowledge of their electronic and optical properties is essential
for their utilization in opto-electronic devices.

It is a well known fact that the standard DFT based approaches normally
underestimate the band gaps of bulk semiconductors because they do
not include the quantum many-body effects\cite{gw_acc}. However,
neglecting many-body effects in low-dimensional systems such as nanoribbons
are likely to cause much more severe errors as compared to bulk systems,
because electron-correlation effects are enhanced in lower dimensions
\cite{GNR_GW,GNR_GW2,GNR_GW3,GNR_GW4,GNR_GW5}. All the aforementioned
electronic structure studies of SiCNRs have been based upon DFT with
various exchange-correlation functionals without accounting for the
many-body effects. Furthermore, it will also be of tremendous interest
to explore the optical absorption in SiCNRs in order to explore the
possibilities of their applications in opto-electronic devices. With
these goals in mind, in this work we present a systematic study of
ASiCNRs in which we go beyond the mean field DFT approach to compute
the influence of many-body effects on their band structure within
the GW approximation. We find significant GW corrections of up to
2 eV to the Kohn-Sham band gaps. Furthermore, we also calculate the
optical absorption spectra of these systems within a Bethe-Salpeter
equation (BSE) based approach, which includes electron-hole interaction,
and thus excitonic effects. The absorption spectra of ASiCNRs are
found to be dominated by strongly bound exciton peaks with binding
energies up to 2eV.  We also performed calculations of formation energies of these nanoribbons for various widths,  and conclude that with the increasing widths nanoribbons become more stable, and, for a given width, hydrogen-satured ribbons are much more stable as compared to the bare ones.

\section{Computational Details}

In earlier works from our group we studied the opto-electronic properties
of graphene nanoribbons using a semi-empirical Hamiltonian based 
methodology\cite{GNR_kond1,GNR_kond2}.
In the present calculations, we have adopted a plane-wave based first-principles
methodology as implemented in the ABINIT package.\cite{abinit1,abinit2}
First, the calculations were performed for the ground state of the nanoribbon concerned, using  
DFT-GGA methodology, employing Perdew, Burke, Enzerhof
(PBE) pseudopotentials.\cite{psp} During the geometry optimization, a kinetic energy cutoff of 26 Hartree (1 Hatree
= 27.21 eV) was used, while for Brillouin zone integration,  a k-point grid with $1\times1\times11$ points was employed. For  total energy/cell, convergence threshold
of $10^{-10}$ Hartree (Ha) was used, and during the structural relaxations,  atomic force cutoff of  $5\times10^{-5}$
Ha/Bohr was employed.  Once the geometry optimization was complete, charge
density profiles, and denisty of states, were computed using the software package
VASP,\cite{VASP} employing the same exchange-correlation potential, and the kinetic energy cutoff as in ABINIT, but  a superior k-point mesh of dimensions $1\times1\times45$. The GGA wave functions
are a good starting point for the many body perturbation theoretic
computations, aimed at including the electron-correlation effects.
The screening effects are included within a plasmon pole approximation,
\cite{ppm} and quasiparticle corrections to the GGA eigenvalues
are then evaluated using the G0W0 approximation.\cite{g0w0}  For the exchange and correlation parts of self energy calculations, cut
off energies of 18 Ha, and 6 Ha, respectively, were used.
 A $1\times1\times30$ k-point
grid ensured convergence of the GW band gap to within 0.1 eV for the
systems considered here. A total of 250 bands was employed for the  GW calculations, out of which  the number of virtual (unoccupied) bands was $>150$. Thus, the choice of computational parameters employed in the GW calculations suggests that these calculations are well converged. 
Lastly, the BSE is solved to obtain the
optical response with the inclusion of the electron-hole interaction,
retaining only the resonant part of the Bethe-Salpeter Hamiltonian,
because the inclusion of the coupling had a negligible effect. In
the GW and the BSE calculations, a Coulomb truncation scheme was also
used to avoid the consideration of long range interactions between
the periodic images. Nanoribbons were taken to be periodic along the
$z$ direction, and to represent an isolated system, vacuum separations
of more than 11 \AA \ were taken between the adjacent unit cells
along both $x$ and $y$ directions.\\

To denote the widths of the nanoribbons, the usual index notation
is used. ASiCNRs are prefixed by the number of dimer (bond) lines,
$N_{a}$, along the width of the ribbon ($N_{a}$-H-ASiCNR is used
for H-passivated, and $N_{a}$-ASiCNR used for bare armchair nanoribbons).
In this work, we have studied ribbons with $4\leq N_{a}\leq12$, which
for H-ASiCNRs corresponds to widths between 0.7 nm and 2.3 nm, and
for bare ASiCNRs to widths  between 0.46 nm and 1.7 nm.

\vspace{0.2in}

\begin{figure}[ht]
\centering \includegraphics{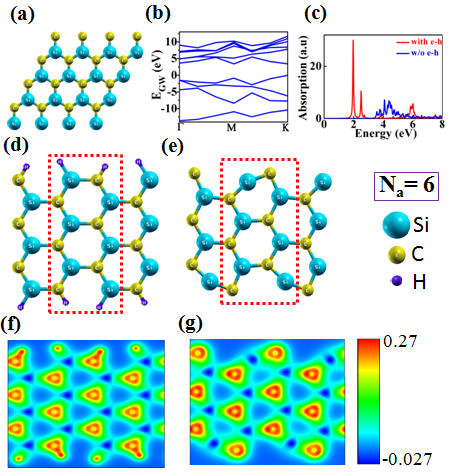}\protect\protect\protect\caption{(Color online) Geometric structures of (a) 2D SiC sheet 
(d) Hydrogen
passivated, (e) bare armchair SiC nanoribbons, . Both types of SiCNRs
correspond to $N_{a}=6$, and unit cells employed for the calculations
are shown in the dashed boxes. Panels (b) and (c) show the calculated GW band
structure and optical absorption spectra of SiC sheet. Panels (f) and (g) display 2D
charge density (in the units of $e$/Bohr$^3$) plots of 6-H-ASiCNR, and 6-ASiCNR, respectively. }

\label{fig:gstructure} 
\end{figure}

\vspace{0.2in}
 .

\section{Results and Discussion}

In this section we present the results of our calculations on the
quasiparticle band structure and the optical absorption spectra of
ASiCNRs. Before that, however, we use the same methodology to compute
the band structure of 2D SiC sheet at the GW level, in order to validate our approach.

\subsection{SiC Monolayer}

Because the calculations presented here are the first reported GW
calculations on ASiCNRs, it is important to benchmark the underlying
approach. Therefore, we first use the same methodology to compute
the quasi-particle band structure of monolayer SiC for which GW calculations
had been reported earlier by Louie and coworkers \cite{sic_bulk_gw}.
The honeycomb structure adopted for the infinite SiC sheet is shown
in  Fig. 1(a), while its calculated GW band structure is presented
in Fig. 1(b). We obtain a direct band gap of 3.4 eV located at the
K point of Brillouin zone, at the GW level, which is 0.9 eV greater
than our DFT-GGA band gap of 2.5 eV. Our GW result is found to be
in very good quantitative agreement with the GW value of 3.6 eV, also
located at the K point, reported by Louie group \cite{sic_bulk_gw}.
Furthermore, we also calculated excitonic binding energy of 2D SiC sheet using the GW+BSE approach.  Fig 1 (c) shows the optical absorption spectra of sheet, with (red curve), and without (blue curve), electron-hole interactions. The difference between the first prominent peaks of these two curves gives the binding energy of 1.5 eV, which is 0.3 eV more than the value 1.2 eV reported by Louie and coworkers\cite{sic_bulk_gw}. This agreement between the GW and 
BSE calculations performed using two different packages, assures us of the 
quality of our calculations.

\subsection{Quasiparticle energies of hydrogenated armchair SiC nanoribbons}

\vspace{0.1in}

\begin{figure}[ht]
\centering \includegraphics{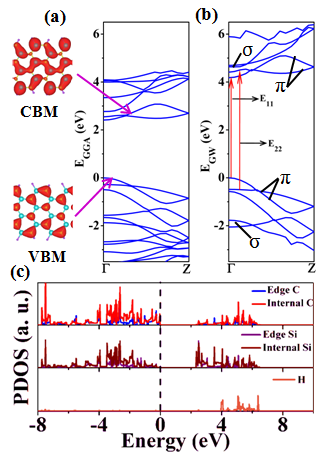}\protect\protect\protect\caption{(Color online) For 6-H-ASiCNR: (a) DFT-GGA band structure (pink arrows show the charge densities corresponding to 
VBM and CBM) (b) GW band structures (red arrows denote interband transitions). $\sigma$/$\pi$
  character of bands is also pointed out. (c) Atom projected density of states  (black dashed line denotes the Fermi level)}

\label{fig:gstructure1} 
\end{figure}

\vspace{0.1in}
 First we present and discuss the results of hydrogen-passivated ASiCNRs
(H-ASiCNRs), for which three different families of width, that is, $N_{a}=3p$,
$N_{a}=3p+1$, and $N_{a}=3p+2$ ($p$ an integer), have been studied,
with $N_{a}$ ranging from 4 to 12. The geometrical
structure of H-ASiCNR for $N_{a}=6$ is shown in Fig. 1(d), for which
our optimized average Si-C bond length is 1.78 \AA , while Si-H and
C-H bond lengths are 1.49 \AA \ and 1.09 \AA ,
respectively. Table 1 presents the band gaps at
the GGA and GW levels, and the exciton binding energies for $N_{a}$=
6, 7, 8. The GW self-energy corrections increase their band gaps by
1.6-1.8 eV, when compared to their GGA values. Thus, the magnitude of
the increase in band gaps for H-ASiCNRs due to electron correlation
effects is significantly greater than the 1 eV increase observed in
the monolayer discussed in the previous section, and for the increase
reported for the bulk SiC \cite{sic_bulk_gw}. The larger correction
observed for nanoribbons is a consequence of enhanced Coulomb interaction
effects, and reduced screening, in lower dimensions. \vspace{0.2in}

\begin{figure}[ht]
\centering \includegraphics{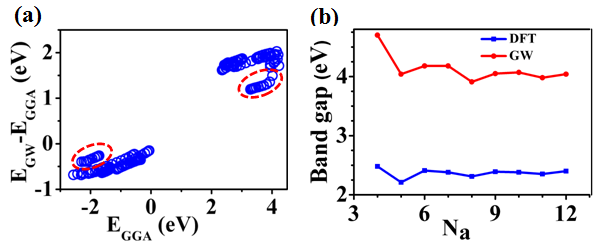} \protect\protect\protect\caption{(Color online) (a) Quasiparticle self-energy corrections to the GGA
Kohn-Sham energies. Corrections to the nearly free electron states
are enclosed by ellipses (red-dashed), and (b) variation of the GGA band gap and quasiparticle
gaps with ribbon width for H-ASiCNRs. }

\label{fig:qpwidth} 
\end{figure}

\vspace{0.2in}

Next, we specifically discuss the case of 6-H-ASiCNR (notation $N_{a}$-H-ASiCNR),
because the other two studied H-ASiCNRs exhibit similar behavior.
In Figs. 2 (a) and 2 (b), we present our calculated GGA and GW quasiparticle
band structures of 6-H-ASiCNR,   from where it is clear
that the hydrogen passivated ASiCNRs are direct band gap semiconductors
as their valence band maximum (VBM) and conduction band minimum (CBM)
occur at the high symmetrical point $\Gamma$ in brillioun zone. Partial
charge denisty plots for VBM and CBM of 6-H-ASiCNRs are shown in left
side panel of Fig 2 (a). These plots reveal that main contribution to the VBM charge density is from
electrons distributed around the carbon atoms, while electrons from 
from both  Si and C atoms contribute to the CBM charge density. Fig. 2 (c) shows the atom projected density
of states (PDOS) of 6-H-ASiCNR.  PDOS has been plotted both for internal
atoms, as well as the ones located on the edges. It clearly show that edge and internal C atoms contribute mainly to the states states near to the
VBM, while internal Si atoms contribute dominantly to form the states close to the
CBM. The contribution of hydrogen atoms appears in the conduction band states far away from
the CBM. The self energy corrections to various bands are plotted
in Fig. 3 (a), and are dependent on their nature. The $\pi$ band
states associated with the $p_{z}$ orbitals are perpendicular to
the ribbon plane, extending into the vacuum region. As a result, the
many body interactions for the $\pi$ states are less screened, and
lead to large quasiparticle corrections. On the other hand, the $\sigma$
band states lie in the plane of the ribbon, and experience greater
screening. Thus, the quasiparticle corrections for the $\sigma$ states
are smaller than the $\pi$ states. In H-ASiCNRs, the bands around
the Fermi energy comprise only of the $\pi$ states, which exhibit
corrections which are mostly linear as a function of the energy. A
few points (encircled, \emph{cf}. Fig. 3a) which deviate from the main
linear region correspond to the nearly free electron states which
are loosely bound to the ribbon, and have weak many-body interactions
resulting in smaller quasiparticle corrections \cite{GNR_GW2}. Compared
to their GGA shapes, the bands not only shift, but also stretch by
an average 30\%--40\%, and even cross due to the non-uniform quasiparticle
corrections resulting from the GW calculations. The family behavior
of the band gaps is preserved at the GW level, as the correction is
roughly proportional to the GGA band gap. Fig. 3(b) shows the band
gap as a function of the ribbon width in GGA and GW approximation
for H-ASiCNRs. From Table 1 it is obvious that the family behaviour of DFT
and GW band gaps follow the heirachy  $3p>3p+1>3p+2$, in good agreement with earlier studies of DFT band gaps of H-ASiCNRs,\cite{ASiCNR_Eg_osc}
and quite unlike H terminated graphene~\cite{GNR_GW} and boron-nitride~\cite{GNR_GW4} nanoribbons which follow the hierarchy  $3p+1>3p>3p+2$\\

\vspace{0.2in}

\begin{table}[h]
\begin{tabular}{ccccc}
\hline 
$N_{a}$  & $E_{GGA}$  & E$_{GW}$  & $\Delta E$  & E$_{b}$\tabularnewline
\hline 
6 ($3p)$  & 2.41  & 4.18  & 1.77  & 1.07\tabularnewline
7 ($3p+1$)  & 2.38  & 4.17  & 1.79  & 1.76\tabularnewline
8 ($3p+2$)  & 2.31  & 3.91  & 1.60  & 1.93\tabularnewline
\hline 
\end{tabular}\protect\protect\protect\caption{The values of GGA and quasiparticle band gaps and their difference
for H-ASiCNRs are tabulated. The last column gives the highest exciton
binding energy. All energy values are in eV.}

\label{t1} 
\end{table}

\subsection{Optical absorption spectra of H-ASiCNRs}

Based on GW quasiparticle band structures, we calculated optical absorption
spectra of H-ASiCNRs, with (GW-BSE), and without (GW-RPA) electron-hole
interaction, and results are presented in Fig. 4. BSE calculations
were performed by including five valence and five conduction bands.
The polarization direction of the radiation is assumed to be along
the length of the ribbons, as significant quenching occurs in 1D systems
for perpendicular directions. The blue colored curves denote the absorption
due to inter-band transitions at the GW-RPA level. The red colored
curve denotes the absorption obtained by solving the Bethe Salpeter
equation, including  both the electron-electron, and electron-hole interactions.

\vspace{0.2in}

\begin{figure}[h]
\centering \includegraphics{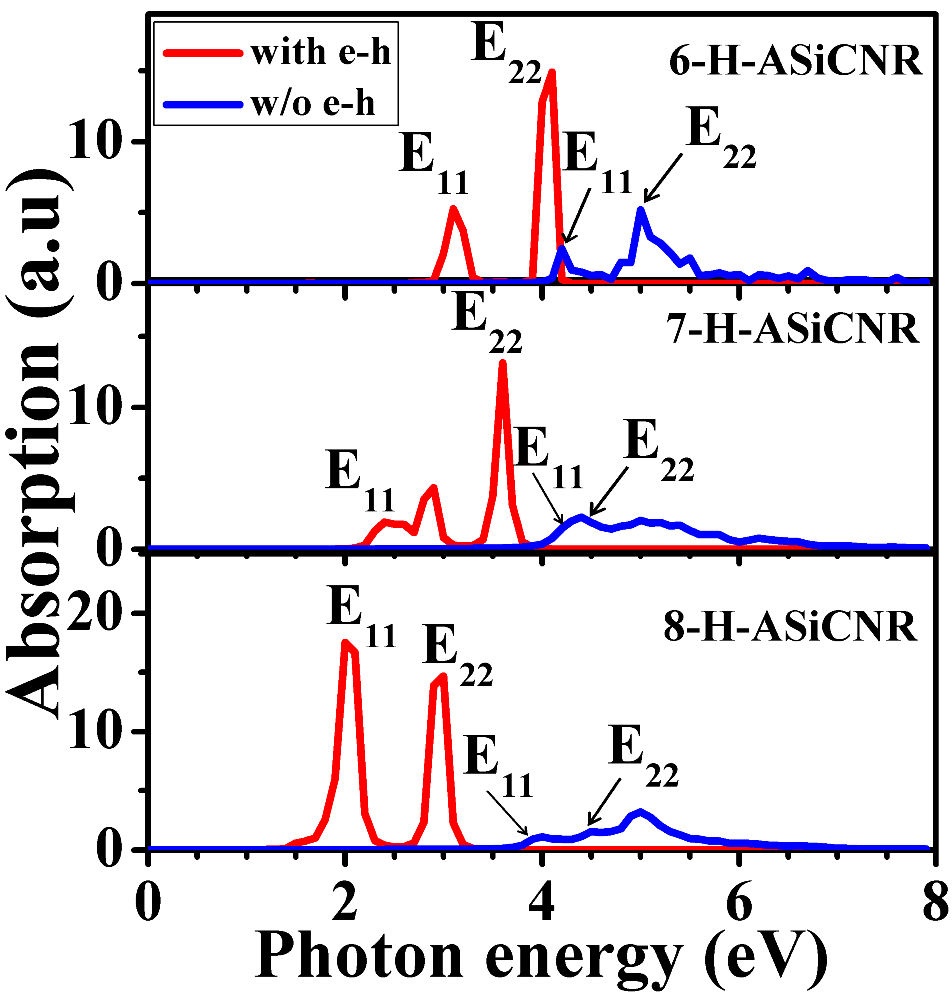} \protect\protect\protect\caption{(Color online) Optical absorption spectra of the H-ASiCNRs, with (red)
and without (blue) electron-hole interaction. Lorentzian broadening
of 0.03 eV has been used.}

\label{fig:bseHanr} 
\end{figure}

\vspace{0.2in}

As compared to the independent particle absorption spectra, BSE based
spectra exhibit changes both in the positions of the peaks, as well
as the line shapes, with excitonic states having binding energies
in the range  1.07--1.93 eV. We adopt the following notation to label interband
transitions: $E_{nm}$ denotes the interband transitions from the
$n$th valence band to the $m$th conduction band, with the band indices
counted from the Fermi level. The underlying band-to-band transitions
are also marked with arrows in the band structure plot in Fig. 2 (b).
The first peak ($E_{11}$) of ribbon arises due transitions between
the highest valence band, and the lowest conduction band at the zone
center, while other prominent peak arises due to transitions between
the second valence and conduction bands. Weak peaks in between include
contributions from cross transitions between these bands. For the
case of $N_{a}=6,7$, $E_{22}$ peak is the most intense one, while
for $N_{a}=8$ case, $E_{11}$ peak acquires significant oscillator
strength, and becomes the most intense peak. The binding energies
of the excitons corresponding to the $E_{11}$ peak are 1.07 eV, 1.76
eV, and 1.93 eV, and for $N_{a}=6,\;7,\;\mbox{and }8$, respectively.
In comparison, our computed value of exciton binding energy of SiC sheet is 1.5 eV, while that of bulk
2H-SiC is reported to be 0.1 eV\cite{sic_bulk_gw}. Thus, exciton binding energy for ASiCNRs is comparable
to that of the monolayer, but significantly larger than that of the
3D bulk SiC. This large difference between the exciton binding energies
of ASiCNRs and bulk SiC can be attributed to their reduced dimensionality,
because in reduced dimensions, electron-hole interactions get stronger
as quantum confinement increases the overlap between electron and
hole wave functions, leading to the formation of small radius excitons\cite{GNR_GW4}. When we compare the exciton binding energies of nanoribbons of different widths, we find a strong width dependence, implying that the 
strength of electron-hole interactions is also strongly width dependent.
This suggests that one can tune the opto-electronic properties of 
H-ASiCNRs, which are direct band gap materials, by manipulating their width, a property highly useful from the point of view of  device applications.

\subsection{Quasiparticle energies of bare armchair SiC nanoribbons}

 In this section, we present and discuss the quasi-particle band structure
of bare ASiCNRs. Fig 1 (e) displays the relaxed geometric structure
of bare 6-ASiCNR, from which it is obvious that it undergoes significant
edge reconstruction because of the dangling bonds, and the Si-C bond
length at the edge is reduced from 1.78 \AA \ to 1.69 \AA . The effect of edge reconstruction is also obvious in total charge density plot of the ground states of bare 6-ASiCNRs (Fig. 1 (g)), 
when compared with that of 6-H-ASiCNR (Fig. 1 (f)). Furthermore, we find
that, at all levels of theory, for the same width, the band gaps of
bare ASiCNRs are smaller as compared to those of H-ASiCNRs.

\begin{figure}[ht]
\centering \includegraphics{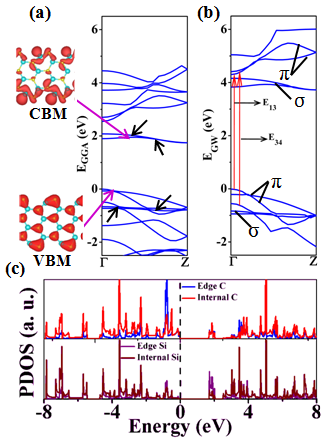} \protect\protect\protect\caption{(Color online) (a) DFT-GGA, and (b) GW band structures of bare 6-ASiCNR.{}
Red color arrows denote the interband transitions, pink color arrows show the VBM and CBM, black color arrows denote the additional valence and conduction bands. $\sigma$/$\pi$
  character of bands is also pointed out. (c) Partial density of states for 6-ASiCNR,
Fermi level shown with black dashed lines.}

\label{fig:g1structure} 
\end{figure}

 At the
DFT-GGA level, they are smaller by $\approx$0.7 eV , a result in
agreement with the results of previous theoretical studies,\cite{sic_bare}
while at the GW level, the difference reduces to $\approx0.3-0.4$
eV. For bare ASiCNRs also we studied the family behavior by considering
 widths of $N_{a}$= 6 ($3p$), 7 ($3p+1$), 8 ($3p+2$);,  the
results of our calculations are presented in the Table \ref{t2}.
 Figure \ref{fig:g1structure} displays the calculated GGA and
GW band structures for $N_{a}=6$. From the figure and the table it
is obvious that, as in the case of hydrogen-passivated ribbons, quasiparticle
corrections due to the GW approach broaden the band gaps of bare ASiCNRs
also by about 2 eV. However, unlike the hydrogen-passivated ribbons,
the bare ASiCNRs exhibit an indirect band gap because the top of the
valence band, and the bottom of the conduction band, are located at
points $\Gamma$ and $Z$ of the Brilliouin zone, respectively, whereas
in H-ASiCNRs both were located at the $\Gamma$ point. 
As shown in Fig. \ref{fig:g1structure}, as compared to H-ASiCNRs, for the case of  bare ASiCNRs  we can observe that two new, relatively flatter bands, appear near both the CBM and the VBM.
As far as the CBM region is concerned, it is now entirely composed of these two new bands, while for the valence band case, the new bands are slightly away from the VBM. This restructuring of the band structure caused by hydrogen
removal leads not only to a reduced band gap, but also transforms bare ASiCNRs into indirect
band gap materials, as compared to H-ASiCNRs. Charge density analysis reveals that these 
additional valence and conduction bands 
are due to the dangling edge bonds, and, because, the electrons involved do
not participate in the bond formation, they have relatively less interaction
with each other, leading to flatter bands.  Partial charge desity plots for VBM and
CBM of bare ASiCNRs are presented in  Fig 5 (a). A comparison with the corresponding plots of H-ASiCNRs reveals that: (a) VBMs of bare ribbons are similar to the case of hydrogen saturated ones in that they derive their charge density contributions from the same set of carbon atoms, and (b) CBMs of bare ribbons are entirely different from those of the saturated ones in that their charged density is almost entirely due to the edge silicon atoms. 
From the PDOS plots of bare ASiCNRs presented in Fig 5 (c), one can deduce that 
the main contribution to the states near VBM is from  interior C atoms, while the contribution to the states near CBM is mainly due to  the edge Si atoms.

\begin{figure}[ht]
\centering \includegraphics{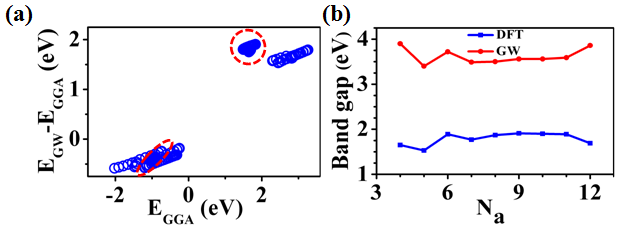} \protect\protect\protect\caption{ For bare ASiCNRs: (a) quasi-particle self-energy corrections to
the GGA
Kohn-Sham energies. Corrections to the states corresponding to
flat bands in  conduction (valence) band regions are enclosed inside red-dashed circle (ellipse).
(b) Variation of the GGA and quasi-particle band gaps as a function of
the ribbon width. }

\label{fig:qp2width} 
\end{figure}

\vspace{0.2in}

In Fig. \ref{fig:qp2width}(a), we show the self energy corrections
for ASiCNRs, as a function of energies. The corrections to the GGA
eigenvalues of bare ribbons are qualitatively similar to those for
H-ASiCNRs, with the corrections to the $\pi$ states being larger,
as compared to those for the $\sigma$ states, and exhibit a linear
variation with respect to the energy. The states enclosed in red circle,
which deviate from the main region, represent corrections to the conduction
band states localized on the edge atoms. In Fig. \ref{fig:qp2width}(b)
we present the variation of the band gap as a function of the ribbon
width, and it is clear that bare ASiCNRs follow the same trends
as H-ASiCNRs in that the band gap exhibits oscillatory behavior. For the largest width 
considered ($N_{a}=12$), GW band gap of H-ASiCNR
is 0.64 eV larger than that of the SiC sheet, while for bare ribbon
the same difference is 0.46 eV.  With the increasing width, band gaps
of  hydrogen passivated nanoribbons are expected
to approach the band gap of the SiC monolayer, while in case of bare ones it is unlikely, 
because of the presence of flat bands (Fig.5(a)). \vspace{0.2in}

\begin{table}[h]
\begin{tabular}{ccccc}
\hline 
N  & $E_{GGA}$  & E$_{GW}$  & $\Delta E$  & E$_{b}$\tabularnewline
\hline 
6 ($3p$)  & 1.74  & 3.72  & 1.98  &  0.62\tabularnewline
7 ($3p+1$)  & 1.51  & 3.48  & 1.97  & 1.26\tabularnewline
8 ($3p+2$)  & 1.62  & 3.56  & 1.94  & 2.48\tabularnewline
\hline 
\end{tabular}\protect\protect\protect\caption{Values of GGA and GW band gaps, and their difference for bare ASiCNRs
are tabulated. The last column gives the highest exciton binding energy
for the given width. All energies are in eV units.}

\label{t2} 
\end{table}

\vspace{0.2in}

\subsection{Optical absorption spectra of bare ASiCNRs}

\begin{figure}[h]
\centering \includegraphics{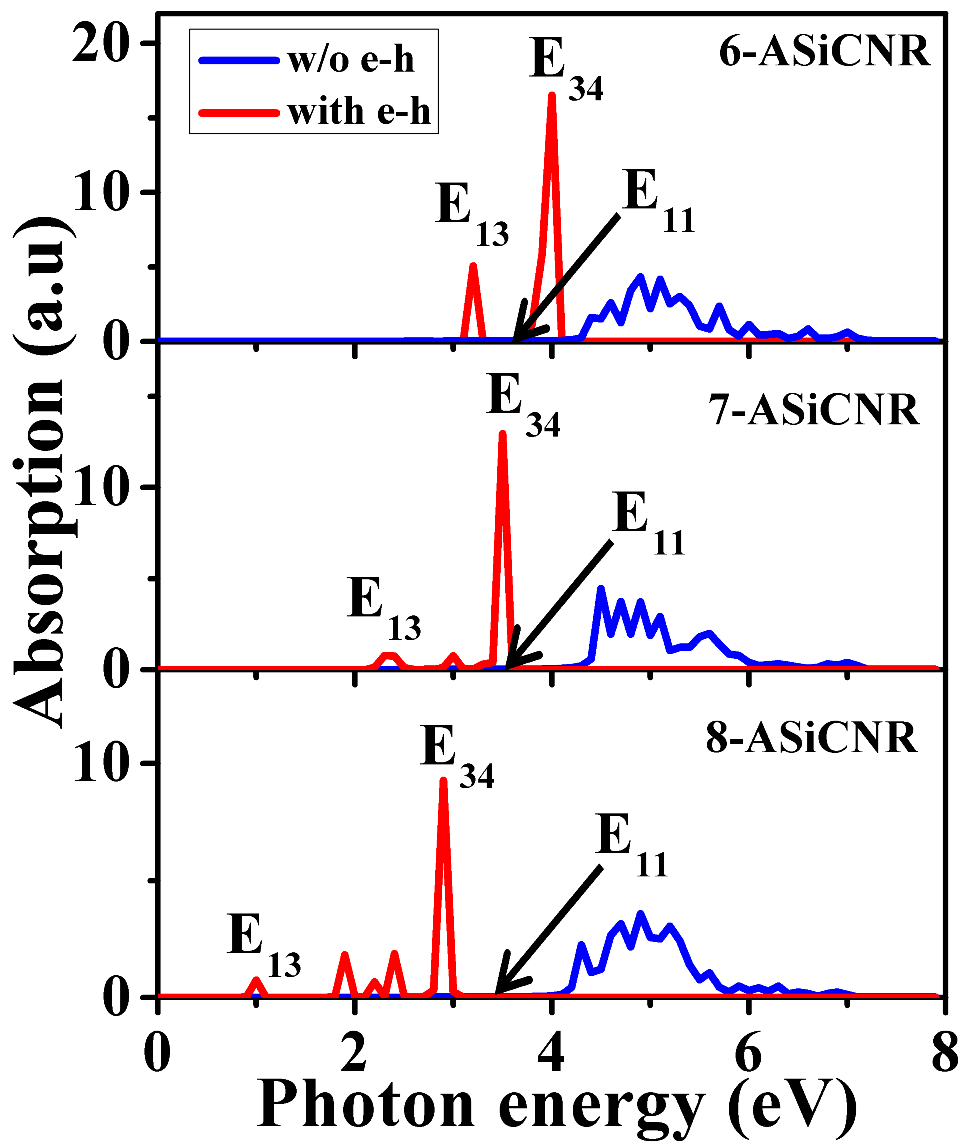} \protect\protect\protect\caption{Optical absorption spectrum of bare 6-ASiCNRs, with (GW-BSE), and
without (GW-RPA), electron-hole interactions. Lorentzian broadening
of 0.03 eV has been used.}

\label{fig:bseanr} 
\end{figure}

In Fig. \ref{fig:bseanr} we present the optical absorption spectrum
of bare ASiCNRs, for $N_{a}=6\mbox{--}8$. For the calculations of
the BSE spectra, five valence and five conduction bands were included.
As for the case of H-ASiCNRs, once the electron-hole interaction effects
are incorporated, the resulting BSE spectrum of bare ribbons looks
completely different as compared to the band-to-band absorption spectrum
computed at the GW-RPA level. Similar to the case of H-ASiCNRs, the
excitons in bare ribbons are also strongly bound, with the binding
energies in bare ribbons, for the same ribbon widths, being larger
than in the hydrogen passivated ones (\emph{cf}. Tables \ref{t1}
and \ref{t2}). We can also find strongly bound excitons in bare ASiCNRs
with binding energies of in the range  0.62--2.45 eV. The reasons behind
these large binding energies are the same as in case of H-ASiCNRs,
i.e., stronger electron-hole interactions due to quantum confinement,
and reduced screening in lower dimensions. \vspace{0.2in}

\vspace{0.2in}
 One qualitative difference in the optical properties of bare ribbons
as compared to the hydrogen-passivated ones is that bare ribbons do
not absorb at the band gap because they exhibit an indirect band gap,
as discussed in the previous section. Therefore, the prominent peaks
displayed in Fig. \ref{fig:bseanr} do not have an $E_{11}$ contribution.
 In all the ribbons considered, the first peak corresponds to an $E_{13}$
transition, while the strongest peak is due to $E_{34}$ transition.
Because of the indirect nature of excitons, bare ASiCNRs cannot be
used in devices such as LEDs, however, they may be useful in transport
oriented applications. As far as the location of first absorption
peak is concerned, both hydrogen-passivated and bare nanoribbons follow
the pattern $E(3p+2)>E(3p+1)>E(3p).$  Because of the presence of an indirect 
band gap in bare ASiCNRs, these materials will have poor photoluminscence efficiency, and thus will not be useful in light-generation applications.

\subsection{Energetic Stability}

Energetic stability of hydrogen terminated ASiCNRs is
well known from previous theoretical studies.\cite{ASiCNR_Eg_osc}
In order to check the stability of bare ASiCNRs, we performed calculations of  their  formation 
energies, and compared them with those of H-ASiCNRs . Following formula was employed for computing the edge formation energy per unit length of edge of  a given nanoribbon,

$$E_{f}=\left(E_c-N_d  E_{2D}-0.5 N_H  E_{H_2} \right)/2L,$$ 

where $E_{c}$ is the total energy/cell of the nanoribbon under consideration, $E_{2D}$ is the total energy/cell of infinite 2D SiC sheet,  $E_{H_{2}}$ is the total energy of a hydrogen molecule, $N_d$ is the total number of SiC units in the unit cell, $N_H$ is the total number of H atoms in the unit cell ($N_H=0$, for bare nanoribbons), while $L$ is the length of the edge (in \AA\, unit). It is easy to see that a lower value of $E_f$ will imply a higher stability 
of the considered nanoribbon.

\begin{figure}[h]
\centering \includegraphics{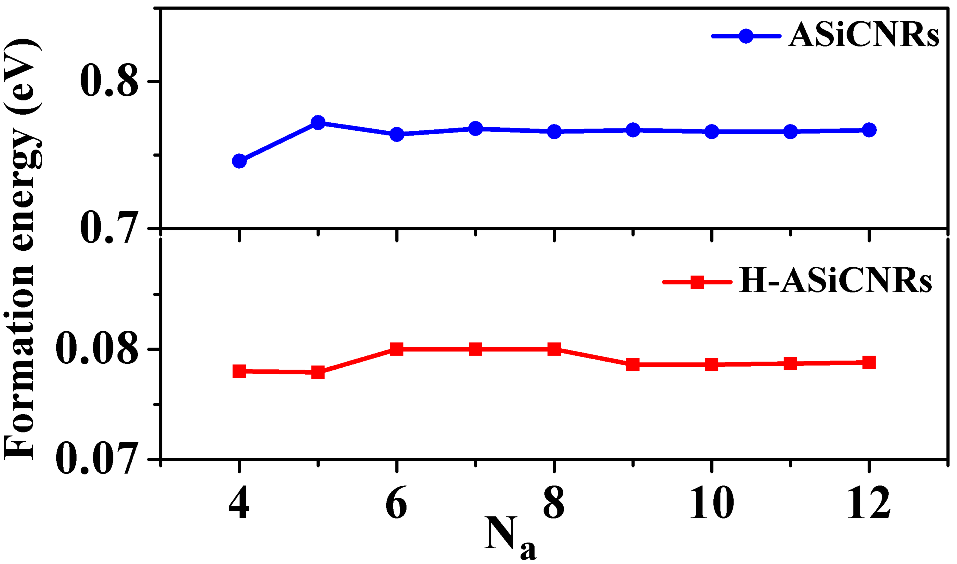} \protect\protect\protect\caption{Formation energies of bare ASiCNRs and H-ASiCNRs as a function of width
}

\label{fig:fe} 
\end{figure}

Formation energy as a function of nanoribbon width for both bare and 
H-saturated nanoribbons studied here is presented in Fig 8. 
The fact that both types of nanoribbons exhibit little change in the value of edge formation energy with the increasing width,  implies a weak  width dependence.  H-ASiCNRs studied here have a much lower average formation energy of 0.07 eV/\AA \ as compared to  0.76 eV/\AA \ of the bare ones,  
clearly demonstrating that H-ASiCNRs  will be much more stable when compared to the bare ones. 
Therefore, based upon the formation energy data, we conclude that the synthesis of H-ASiCNRs in 
laboratory will be much more favourable as compared to the bare ones.

\section{Conclusions}

To summarize, we performed large-scale first-principles calculations
on armchair-type SiC nanoribbons of widths in the range $4\leq N_{a}\leq12$,
in order to study their electronic structure and optical properties.
Both hydrogen passivated and bare edge configurations were considered,
 sophisticated many-body methods such as the GW approximation for
the band structure, and BSE for the optical absorption, were adopted,
and the convergence of the calculations was carefully checked. Therefore,
the results presented here are fairly accurate, and constitute, to
the best of our knowledge, the first set of GW and BSE calculations for
SiCNRs.  From our calculations we
conclude that hydrogen-passivated ribbons are direct band gap semiconductors,
and bare edge-ribbons undergo significant edge reconstruction, and  become
indirect band gap semiconductors.
As expected, self-energy corrections widened the
band gaps,  and excitonic effects modified optical absorption
spectra dramatically for both the H-saturated and the bare ribbons. Our results predict that quasiparticle
band gaps of hydrogen passivated and bare ASiCNRs are $\approx$ 2 eV larger
than their DFT-GGA band gaps, due to the inclusion of many-body effects within the GW approximation. These large quasiparticle corrections to the band gaps suggest enhanced Coulomb correlation effects in reduced dimensions. Our BSE based calculations of the excitonic effects predict  large excitonic binding energies in the range of 0.62--2.45 eV
in optical absorption spectra of both types of ASiCNRs, which is yet another consequence of reduced dimensions. Key results of our calculations can be tested in future experiments once these 
nanoribbons are synthesized in laboratories. Theoretically speaking, it will 
also be of interest to explore other ribbon geometries of SiC, and also ribbons with unequal 
proportions of Si and C.

\section*{Acknowledgements}

All authors gratefully acknowledge computaional support from Monash
Sun Grid cluster and the National Computing Infrastucture, Australia.
N.A. gratefully thanks P. Chan and S. Ayyub for helping in allocating
the required CPUs to carry out the calculations. 

\vspace*{0.5cm}

\bibliographystyle{iopart-num}
\bibliography{sicanr}

\providecommand{\newblock}{}
\begin{thebibliography}{10}
\expandafter\ifx\csname url\endcsname\relax
  \def\url#1{{\tt #1}}\fi
\expandafter\ifx\csname urlprefix\endcsname\relax\def\urlprefix{URL }\fi
\providecommand{\eprint}[2][]{\url{#2}}

\bibitem{graphene_synth}
Novoselov K~S, Geim A~K, Morozov S~V, Jiang D, Zhang Y, Dubonos S~V, Grigorieva
  I~V and Firsov A~A 2004 {\em Science\/} {\bf 306} 666--669 (\textit{Preprint}
  \eprint{http://www.sciencemag.org/content/306/5696/666.full.pdf})
  \urlprefix\url{http://www.sciencemag.org/content/306/5696/666.abstract}

\bibitem{gn1}
Geim A~K and Novoselov K~S 2007 {\em Nature materials\/} {\bf 6} 183--191

\bibitem{gn2}
Katsnelson M~I 2007 {\em Materials today\/} {\bf 10} 20--27

\bibitem{gn3}
Castro~Neto A~H, Guinea F, Peres N~M~R, Novoselov K~S and Geim A~K 2009 {\em
  Rev. Mod. Phys.\/} {\bf 81}(1) 109--162
  \urlprefix\url{http://link.aps.org/doi/10.1103/RevModPhys.81.109}

\bibitem{bn_synth}
Coleman J~N, Lotya M, {\'O}Neill A, Bergin S~D, King P~J, Khan U, Young K,
  Gaucher A, De S, Smith R~J {\em et~al.\/} 2011 {\em Science\/} {\bf 331}
  568--571

\bibitem{bn1}
Chopra N~G, Luyken R~J, Cherrey K, Crespi V~H, Cohen M~L, Louie S~G and Zettl A
  1995 {\em Science\/} {\bf 269} 966--967 (\textit{Preprint}
  \eprint{http://www.sciencemag.org/content/269/5226/966.full.pdf})
  \urlprefix\url{http://www.sciencemag.org/content/269/5226/966.abstract}

\bibitem{bn2}
Lee R~S, Gavillet J, Chapelle M~L~d~l, Loiseau A, Cochon J~L, Pigache D,
  Thibault J and Willaime F 2001 {\em Phys. Rev. B\/} {\bf 64}(12) 121405
  \urlprefix\url{http://link.aps.org/doi/10.1103/PhysRevB.64.121405}

\bibitem{bn3}
Park C~H and Louie S~G 2008 {\em Nano Letters\/} {\bf 8} 2200--2203
  (\textit{Preprint} \eprint{http://pubs.acs.org/doi/pdf/10.1021/nl080695i})
  \urlprefix\url{http://pubs.acs.org/doi/abs/10.1021/nl080695i}

\bibitem{bn4}
Guo G~Y and Lin J~C 2005 {\em Phys. Rev. B\/} {\bf 72}(7) 075416
  \urlprefix\url{http://link.aps.org/doi/10.1103/PhysRevB.72.075416}

\bibitem{si_t}
Ni Z, Liu Q, Tang K, Zheng J, Zhou J, Qin R, Gao Z, Yu D and Lu J 2011 {\em
  Nano letters\/} {\bf 12} 113--118

\bibitem{si1}
Fleurence A, Friedlein R, Ozaki T, Kawai H, Wang Y and Yamada-Takamura Y 2012
  {\em Phys. Rev. Lett.\/} {\bf 108}(24) 245501
  \urlprefix\url{http://link.aps.org/doi/10.1103/PhysRevLett.108.245501}

\bibitem{si2}
Feng B, Ding Z, Meng S, Yao Y, He X, Cheng P, Chen L and Wu K 2012 {\em Nano
  Letters\/} {\bf 12} 3507--3511 pMID: 22658061 (\textit{Preprint}
  \eprint{http://dx.doi.org/10.1021/nl301047g})
  \urlprefix\url{http://dx.doi.org/10.1021/nl301047g}

\bibitem{graphitic_SiC}
Miyamoto Y and Yu B~D 2002 {\em Applied physics letters\/} {\bf 80} 586--588

\bibitem{sicmix1}
Menon M, Richter E, Mavrandonakis A, Froudakis G and Andriotis A~N 2004 {\em
  Phys. Rev. B\/} {\bf 69}(11) 115322
  \urlprefix\url{http://link.aps.org/doi/10.1103/PhysRevB.69.115322}

\bibitem{sicmix2}
Mavrandonakis A, Froudakis G~E, Schnell M and M{\"u}hlh{\"a}user M 2003 {\em
  Nano Letters\/} {\bf 3} 1481--1484 (\textit{Preprint}
  \eprint{http://pubs.acs.org/doi/pdf/10.1021/nl0343250})
  \urlprefix\url{http://pubs.acs.org/doi/abs/10.1021/nl0343250}

\bibitem{SiC_mono}
Lin X, Lin S, Xu Y, Hakro A~A, Hasan T, Zhang B, Yu B, Luo J, Li E and Chen H
  2013 {\em Journal of Materials Chemistry C\/} {\bf 1} 2131--2135

\bibitem{ASiCNR_Eg_osc}
Sun L, Li Y, Li Z, Li Q, Zhou Z, Chen Z, Yang J and Hou J 2008 {\em The Journal
  of chemical physics\/} {\bf 129} 174114

\bibitem{sic_bare}
Bekaroglu E, Topsakal M, Cahangirov S and Ciraci S 2010 {\em Phys. Rev. B\/}
  {\bf 81}(7) 075433
  \urlprefix\url{http://link.aps.org/doi/10.1103/PhysRevB.81.075433}

\bibitem{sic2d_eg}
Lu T~Y, Liao X~X, Wang H~Q and Zheng J~C 2012 {\em J. Mater. Chem.\/} {\bf
  22}(19) 10062--10068 \urlprefix\url{http://dx.doi.org/10.1039/C2JM30915G}

\bibitem{SiX}
Ding Y and Wang Y 2013 {\em The Journal of Physical Chemistry C\/} {\bf 117}
  18266--18278

\bibitem{sic_bulk_gw}
Hsueh H~C, Guo G~Y and Louie S~G 2011 {\em Phys. Rev. B\/} {\bf 84}(8) 085404
  \urlprefix\url{http://link.aps.org/doi/10.1103/PhysRevB.84.085404}

\bibitem{SiC_BN_BeO}
Baumeier B, Kr\"uger P and Pollmann J 2007 {\em Phys. Rev. B\/} {\bf 76}(8)
  085407 \urlprefix\url{http://link.aps.org/doi/10.1103/PhysRevB.76.085407}

\bibitem{group4}
\ifmmode~\mbox{\c{S}}\else \c{S}\fi{}ahin H, Cahangirov S, Topsakal M,
  Bekaroglu E, Akturk E, Senger R~T and Ciraci S 2009 {\em Phys. Rev. B\/} {\bf
  80}(15) 155453
  \urlprefix\url{http://link.aps.org/doi/10.1103/PhysRevB.80.155453}

\bibitem{opt_SiCNT}
Wu I~J and Guo G~Y 2007 {\em Phys. Rev. B\/} {\bf 76}(3) 035343
  \urlprefix\url{http://link.aps.org/doi/10.1103/PhysRevB.76.035343}

\bibitem{tight_bind}
Zhao K, Zhao M, Wang Z and Fan Y 2010 {\em Physica E: Low-dimensional Systems
  and Nanostructures\/} {\bf 43} 440 -- 445 ISSN 1386-9477
  \urlprefix\url{http://www.sciencedirect.com/science/article/pii/S1386947710004972}

\bibitem{SiC2D}
Lin S~S 2012 {\em The Journal of Physical Chemistry C\/} {\bf 116} 3951--3955
  (\textit{Preprint} \eprint{http://pubs.acs.org/doi/pdf/10.1021/jp210536m})
  \urlprefix\url{http://pubs.acs.org/doi/abs/10.1021/jp210536m}

\bibitem{sicnr_micro}
Yushin G~N, Cambaz Z~G, Gogotsi Y, Vyshnyakova K~L and Pereselentseva L~N 2008
  {\em Journal of the American Ceramic Society\/} {\bf 91} 83--87 ISSN
  1551-2916 \urlprefix\url{http://dx.doi.org/10.1111/j.1551-2916.2007.02093.x}

\bibitem{3C_nanobelts}
Xi G, Peng Y, Wan S, Li T, Yu W and Qian Y 2004 {\em The Journal of Physical
  Chemistry B\/} {\bf 108} 20102--20104

\bibitem{bicrystaline_nanobelt}
Wu R, Wu L, Yang G, Pan Y, Chen J, Zhai R and Lin J 2007 {\em Journal of
  Physics D: Applied Physics\/} {\bf 40} 3697
  \urlprefix\url{http://stacks.iop.org/0022-3727/40/i=12/a=023}

\bibitem{SiCNR_synth}
Zhang H, Ding W, He K and Li M 2010 {\em Nanoscale research letters\/} {\bf 5}
  1264--1271

\bibitem{SiCNT_synth}
Sun X~H, Li C~P, Wong W~K, Wong N~B, Lee C~S, Lee S~T and Teo B~K 2002 {\em
  Journal of the American Chemical Society\/} {\bf 124} 14464--14471

\bibitem{SiCNT_stable}
Alfieri G and Kimoto T 2014 {\em Applied Physics Letters\/} {\bf 104} 033107

\bibitem{ZSiCNR_DFT}
Lou P and Lee J~Y 2009 {\em The Journal of Physical Chemistry C\/} {\bf 113}
  12637--12640 (\textit{Preprint}
  \eprint{http://pubs.acs.org/doi/pdf/10.1021/jp903155r})
  \urlprefix\url{http://pubs.acs.org/doi/abs/10.1021/jp903155r}

\bibitem{ZSiCNR_Efield}
Lou P and Lee J~Y 2009 {\em The Journal of Physical Chemistry C\/} {\bf 113}
  21213--21217 (\textit{Preprint}
  \eprint{http://pubs.acs.org/doi/pdf/10.1021/jp906558y})
  \urlprefix\url{http://pubs.acs.org/doi/abs/10.1021/jp906558y}

\bibitem{assymetry_H}
Ding Y and Wang Y 2012 {\em Applied Physics Letters\/} {\bf 101} 013102
  \urlprefix\url{http://scitation.aip.org/content/aip/journal/apl/101/1/10.1063/1.4732786}

\bibitem{edge_H_zig}
Lou P 2013 {\em J. Mater. Chem. C\/} {\bf 1}(17) 2996--3003
  \urlprefix\url{http://dx.doi.org/10.1039/C3TC30173G}

\bibitem{O_S_zig}
Lopez-Bezanilla A, Huang J, Kent P~R~C and Sumpter B~G 2013 {\em The Journal of
  Physical Chemistry C\/} {\bf 117} 15447--15455 (\textit{Preprint}
  \eprint{http://dx.doi.org/10.1021/jp406547a})
  \urlprefix\url{http://dx.doi.org/10.1021/jp406547a}

\bibitem{mod_eg}
Zheng F~L, Zhang Y, Zhang J~M and Xu K~W 2011 {\em physica status solidi (b)\/}
  {\bf 248} 1676--1681 ISSN 1521-3951
  \urlprefix\url{http://dx.doi.org/10.1002/pssb.201147001}

\bibitem{B_N_SiCNR}
Costa C and Morbec J 2011 {\em Journal of Physics: Condensed Matter\/} {\bf 23}
  205504

\bibitem{half_bare_zig}
Tang Z~K, Wang L~L, Tang L~M, Huang W~Q, Zhang D~Y, Xu L and Li X~F 2013 {\em
  Solid State Communications\/} {\bf 158} 25 -- 28 ISSN 0038-1098
  \urlprefix\url{http://www.sciencedirect.com/science/article/pii/S0038109813000203}

\bibitem{vac_SiCNR}
Morbec J~M and Rahman G 2013 {\em Phys. Rev. B\/} {\bf 87}(11) 115428
  \urlprefix\url{http://link.aps.org/doi/10.1103/PhysRevB.87.115428}

\bibitem{AGNR_Eg_osc}
Son Y~W, Cohen M~L and Louie S~G 2006 {\em Phys. Rev. Lett.\/} {\bf 97}(21)
  216803 \urlprefix\url{http://link.aps.org/doi/10.1103/PhysRevLett.97.216803}

\bibitem{ABNNR_Eg_osc}
Zhang Z and Guo W 2008 {\em Phys. Rev. B\/} {\bf 77}(7) 075403
  \urlprefix\url{http://link.aps.org/doi/10.1103/PhysRevB.77.075403}

\bibitem{ZGNR_Efield}
Son Y~W, Cohen M~L and Louie S~G 2006 {\em Nature\/} {\bf 444} 347--349

\bibitem{gw_acc}
Godby R~W, Schl\"uter M and Sham L~J 1986 {\em Phys. Rev. Lett.\/} {\bf 56}(22)
  2415--2418
  \urlprefix\url{http://link.aps.org/doi/10.1103/PhysRevLett.56.2415}

\bibitem{GNR_GW}
Prezzi D, Varsano D, Ruini A, Marini A and Molinari E 2008 {\em Phys. Rev. B\/}
  {\bf 77}(4) 041404
  \urlprefix\url{http://link.aps.org/doi/10.1103/PhysRevB.77.041404}

\bibitem{GNR_GW2}
Yang L, Cohen M~L and Louie S~G 2007 {\em Nano Letters\/} {\bf 7} 3112--3115
  (\textit{Preprint} \eprint{http://pubs.acs.org/doi/pdf/10.1021/nl0716404})
  \urlprefix\url{http://pubs.acs.org/doi/abs/10.1021/nl0716404}

\bibitem{GNR_GW3}
Yang L, Cohen M~L and Louie S~G 2008 {\em Phys. Rev. Lett.\/} {\bf 101}(18)
  186401 \urlprefix\url{http://link.aps.org/doi/10.1103/PhysRevLett.101.186401}

\bibitem{GNR_GW4}
Wang S, Chen Q and Wang J 2011 {\em Applied Physics Letters\/} {\bf 99} 063114

\bibitem{GNR_GW5}
Hu C, Ogura R, Onoda N, Konabe S and Watanabe K 2012 {\em Phys. Rev. B\/} {\bf
  85}(24) 245420
  \urlprefix\url{http://link.aps.org/doi/10.1103/PhysRevB.85.245420}

\bibitem{GNR_kond1}
Gundra K and Shukla A 2011 {\em Physical Review B\/} {\bf 83} 075413

\bibitem{GNR_kond2}
Gundra K and Shukla A 2011 {\em Physical Review B\/} {\bf 84} 075442

\bibitem{abinit1}
Gonze X, Amadon B, Anglade P~M, Beuken J~M, Bottin F, Boulanger P, Bruneval F,
  Caliste D, Caracas R, Côté M, Deutsch T, Genovese L, Ghosez P, Giantomassi
  M, Goedecker S, Hamann D, Hermet P, Jollet F, Jomard G, Leroux S, Mancini M,
  Mazevet S, Oliveira M, Onida G, Pouillon Y, Rangel T, Rignanese G~M, Sangalli
  D, Shaltaf R, Torrent M, Verstraete M, Zerah G and Zwanziger J 2009 {\em
  Computer Physics Communications\/} {\bf 180} 2582 -- 2615 ISSN 0010-4655
  \urlprefix\url{http://www.sciencedirect.com/science/article/pii/S0010465509002276}

\bibitem{abinit2}
Gonze X 2005 {\em Zeitschrift f{\"u}r Kristallographie\/} {\bf 220} 558--562

\bibitem{psp}
Fuchs M and Scheffler M 1999 {\em Computer Physics Communications\/} {\bf 119}
  67--98

\bibitem{VASP}
Kresse G and Furthmüller J 1996 {\em Computational Materials Science\/} {\bf
  6} 15 -- 50 ISSN 0927-0256
  \urlprefix\url{http://www.sciencedirect.com/science/article/pii/0927025696000080}

\bibitem{ppm}
Godby R~W and Needs R~J 1989 {\em Phys. Rev. Lett.\/} {\bf 62}(10) 1169--1172
  \urlprefix\url{http://link.aps.org/doi/10.1103/PhysRevLett.62.1169}

\bibitem{g0w0}
Onida G, Reining L and Rubio A 2002 {\em Rev. Mod. Phys.\/} {\bf 74}(2)
  601--659 \urlprefix\url{http://link.aps.org/doi/10.1103/RevModPhys.74.601}

\end{thebibliography}

\end{document}